\documentclass[aps,prd,superscriptaddress,amssymb,amsmath,nofootinbib,showpacs,balancelastpage]{revtex4}
\usepackage{amsmath}
\usepackage{graphicx}
\begin{document}

\title{Thermodynamics of the unified dark fluid with fast
transition}
\author{Ninfa Radicella}
\email{ninfa.radicella@sa.infn.it} \affiliation{Department of
Physics ``E.R. Caianiello'', University of Salerno and INFN,
\\sez. di Napoli, GC di Salerno, Via Giovanni Paolo II, 132, 84084
Fisciano, Salerno, Italy}

\author{Diego Pav\'{o}n}
\email{diego.pavon@uab.es}
\affiliation {Department of Physics, Autonomous University of Barcelona,\\
08193 Bellaterra, Barcelona, Spain}

\begin{abstract}
 In the so-called unified dark fluid models,  the dark
sector gets simplified  because dark matter and dark energy are
replaced by a single fluid that behaves as the former at early
times and as the latter at late times. In this short paper  we
analyze this class of models from the thermodynamic viewpoint.
While the second law of thermodynamics is satisfied, the first two
derivatives of the entropies of the apparent horizon and of the
energy components suffer such a sharp oscillation that doubts are
raised about the soundness of this class of models.
\end{abstract}


\maketitle

 Homogeneous and isotropic world models, aimed at
accounting both for the present era of cosmic accelerated
expansion and the large scale structure, usually assume two dark
components as the chief sources of the gravitational field: dark
matter and dark energy. Combined, they contribute to about  95\%
of the current energy density; the remaining 5\% is provided by
baryons. Very often, both main components are assumed to interact
between them and with the other ingredients of the cosmic budget
(baryons, photons, etc) only gravitationally. The former
characterizes for being cold and, therefore, responsible
(alongside the baryon fluid)  for the formation of galaxies and
clusters thereof. The latter, distinguishes itself for possessing
a huge negative pressure (of the order of its energy density) that
drives the accelerated expansion, and for clustering very weakly
-possibly at the largest accessible scales only. The first one
dominates the expansion at early times ($z > 1$); the latter, at
late times (at redshifts below unity). So, in this scenario, each
dark component has a well-defined and separate role in shaping our
present Universe.
\\  \

\noindent Nevertheless, the possibility was raised that both
components are simply manifestations of a single entity that at
high redshifts would behave as a pressureless fluid and at low
redshifts as a cosmological constant. Well-known examples are the
Chaplygin gas model \cite{kamenschik} and its generalizations, see
e.g. \cite{generalizations}. In these, the cosmic equation of
state (EoS), i.e., the ratio between the pressure and the energy
density, $ w = p/\rho$, gently evolves from zero (the $w$ value
corresponding to cold matter) to $-1$, typical of the quantum
vacuum. In principle, this is an attractive scenario because it
kills two birds with a single stone. Regrettably, it fails at the
perturbative level as the corresponding matter power spectrum is
at variance with observation \cite{sandvik04} and the integrated
Sachs-Wolfe effect significantly departs from the one predicted by
the concordance $\Lambda$CDM  model -see however
\cite{jcap-gorini}. In a related class of models -collectively
known as ``unified dark fluid" (UDF) models- that apparently are
not afflicted by these problems, the transition from the
Einstein-de Sitter regime to the accelerated phase occurs rather
quickly; see \cite{oliver10a,oliver10b,bruni13,yang13} and
references therein. Since in the accelerated phase cosmic
structures essentially stops growing, UDF models feature a longer
matter dominated era  than the $\Lambda$CDM model and unified
models with a slow transition. This may help tell apart both class
of models.
\\  \

\noindent Obviously, viable cosmological models, in addition to
passing the observational tests, must not run into conflict with
well-known physics. The target of this brief paper is to explore
whether UDF models, based on general relativity,  comply with the
second law of thermodynamics. The latter, when applied to systems
that present a horizon (as is the case of the said models),  must
take into account the entropy of matter and fields within the
horizon plus the entropy of the horizon itself, which is
proportional to its area. It arises in a natural way because the
horizon prevents the observer from seeing what lies beyond it.
Here, for the sake of conciseness, we shall focus on the model of
Ref. \cite{yang13} which, we believe, summarizes fairly well the
class of UDF models.
\\ \

\noindent As the cosmological horizon we shall take the apparent
horizon, the marginally trapped surface with vanishing expansion,
since -by contrast to other possible choices- the laws of
thermodynamics are fulfilled on it \cite{bye}. Its radius and
entropy are given by \cite{bak-rey} $\, \tilde{r}_{H} = (H^{2}+k
a^{-2})^{-1/2} \, $ and
\begin{equation}
S_{H} = \frac{k_{B}}{\ell^{2}_{pl}}\, \frac{ \pi}{H^{2} \, + \, k
\, a^{-2}} \, , \label{SH}
\end{equation}
respectively. Here, and throughout, $ \ell_{pl} $ and $k$ denote
the Planck's length and the spatial curvature index.
\\  \

\noindent The second law of thermodynamics simply formalizes the
empirical fact that macroscopic systems spontaneously tend to
thermodynamic equilibrium. In essence it asserts that the entropy,
$S \,$ of isolated systems can never decrease (i.e., $S' \geq 0$),
and that eventually it tends to a maximum (i.e., $ S'' \leq 0$)
compatible with the constraints of the system \cite{callen}. Here
the prime means derivative with respect to the relevant
variable.\footnote{Sometimes one come across  mutilated versions
of this law that leave aside the above condition on $S''\,$. While
it works well for many practical purposes, it is insufficient in
general. Otherwise, one would witness systems with an always
increasing entropy but never achieving equilibrium, something in
sharp contrast with daily experience.}
\\  \

\noindent  Before going any further, we  remark that given the
strong connection between gravity and thermodynamics \cite{jakob,
steven, jacobson, paddy} it is natural to expect that the Universe
behaves as a normal thermodynamic system;  i.e., that it
approaches a state of maximum entropy in the long run \cite{grg1,
grg2}.
\\  \

\noindent UDF models usually enter the following energy
components: radiation, baryons and the unified fluid (subscripts
$b$, $\, r$  and $\, u$, respectively). As mentioned above, the
latter plays the role of cold dark matter at early times and dark
energy later on. Thus, the entropy of the Universe is contributed
to by  the entropy of these plus that of the horizon,
\begin{equation}
S = S_{r} \, + \, S_{b} \, + S_{u} \, + \, S_{H} \, .
\label{Stotal}
\end{equation}
It must never decrease and it must be concave ($S'' < 0$) when $a
\rightarrow \infty$.
\\  \

\noindent On the other hand, the Einstein field equations,
assuming a spatially flat metric metric ($k = 0$), read
\begin{equation}\label{Friedmann}
H^2=H_0^2\left[\frac{\Omega_{b0}}{a^3}+\frac{\Omega_{r0}}{a^4}+\Omega_{u}\right],
\end{equation}
and
\begin{equation}
\frac{H'}{H^2}=-\frac{3}{2\ a H}\left(1+\sum_i
w_i\Omega_i\right)\, , \qquad (i = b, r, u)
\label{Hprime}
\end{equation}
where $w_{r} = 1/3\, $ and $\, w_{b} = 0$. As usual,  the
$\Omega_{i}$ quantities  denote the fractional density
($\Omega_{i} = \rho_{i}/\sum_{i} \rho_{i}$) of the corresponding
component, and a zero subindex attached to a quantity indicates
that it is to be  evaluated at the present time.
\\ \

\noindent Here we consider a UDF model in which the  the
transition is parametrized by the EoS  \cite{bruni13,yang13}
\begin{equation}
w_{u}=-\frac{1}{2}\, \left[\tanh{\left(\frac{a-a_t}{\beta}\right)}
\, + \, 1 \right] \, .
\label{eos}
\end{equation}
It contains two positive-definite, but otherwise free parameters:
$a_{t}$, the scale factor at which the transition takes place, and
$\beta$ that gauges how quickly the transition proceeds (the
smaller $\beta$, the faster the transition). For $a \ll a_{t}$\,
one has $0 \geq w_{u} \gg -1 \,$ , as illustrated in Fig. 1 of
\cite{yang13}; in fact, the faster the transition, the smaller the
ratio $-a_t/\beta$ and the hyperbolic tangent approaches $-1$. On
the other side, when $a \gg a_{t}$, the hyperbolic tangent tends
to $1$ and $w_{u}\simeq-1$.
\\  \

\begin{figure}[htb]
\centering
\includegraphics[width=10cm]{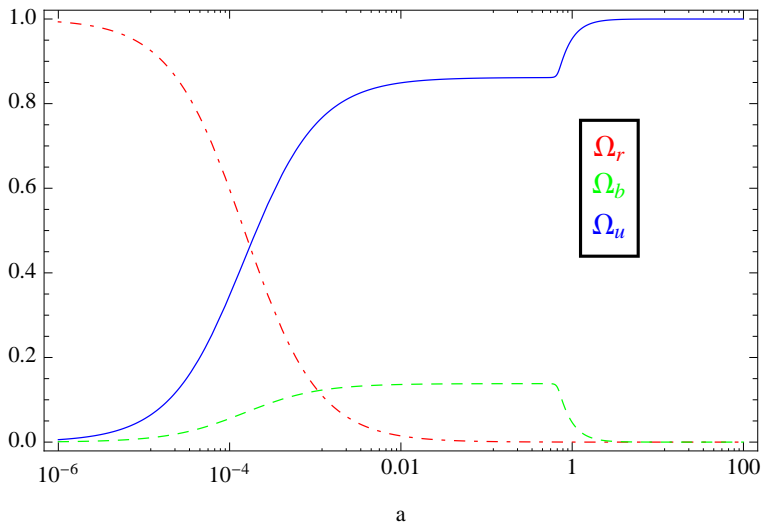}
\caption{{\small Evolution of the density parameters $\Omega_{i}$
of the UDF fast transition model of Ref. \cite{yang13}. Solid,
dot-dashed, and dashed lines denote $\Omega_{u},\, \Omega_{r}$ and
$\Omega_{b}$, respectively. In plotting this we used the best fit
values derived in \cite{yang13}; namely: $\Omega_{r0}=5 \times
10^{-5}$, $\Omega_{b0}=0.0465$, $H_{0} = 2 \times 10^{-18}\
\text{s}^{-1}$, $a_{t} =0.674$ and $\beta = 0.249$.}}
\label{fig:densities}
\end{figure}

\noindent The conservation equation for the unified fluid, $\,
\rho'_{u} = - 3 (1+w_{u})/a$, leads, after integration, to
\begin{equation}\label{udf}
\Omega_{u}(a)= (1-\Omega_{b0}-\Omega_{r0})\, \exp \left\{- 3\,
{\int_{0}^{a} \frac{[1+w_{u}(x)]}{x} dx}\right\} \, ,
\end{equation}
where, without loss of generality, we have set $a_{0}$ to unity.
Figure \ref{fig:densities} plots the evolution of the fractional
densities of all components (baryons, radiation, and the dark
fluid) using the  best fit values found in \cite{yang13}. The
sudden increase (decrease) of $\, \Omega_{u} \, $ ($\Omega_{b}$)
at $\, a_{t} \sim 0.6 \,$ is a distinguishing feature of the
model.

\begin{figure}[htb]
\centering
\includegraphics[width=10cm]{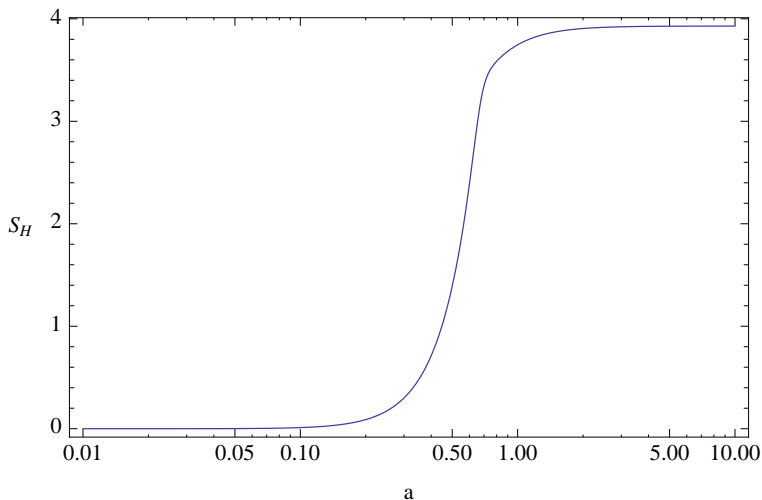}
\caption{{\small Evolution of the entropy of the apparent horizon
in terms of the scale factor.  In plotting the graph we used the
same values employed in the previous figure. The numerical values
in the vertical axis are to be multiplied by the factor $k_{B} \,
\ell_{pl}^{-2} \simeq 5.4 \times 10^{49}$ erg/Kelvin.}}
\label{fig:entropyhor}
\end{figure}

\noindent The evolution of the entropy of the apparent horizon is
shown in Fig. \ref{fig:entropyhor}.
In conformity with the expression
\begin{equation}\label{S'hor}
S'_{H} =\frac{dS_H}{da}\propto \frac{3}{aH^2}\left(1+\sum_i
w_i\Omega_i\right) \geq 0 \,
\end{equation}
[which was obtained with the help of (\ref{Hprime})], it never
diminishes. On the other hand,  the curvature of the graph changes
from positive to negative values  about the transition scale
factor, $a_t$; i.e., when the dark fluid begins to fully dominate
the expansion.
\\  \

\noindent To get the second derivative of the entropy of the
apparent horizon we first express the derivative of the fractional
densities of the various components in terms of the original
quantities, i.e.,
\begin{equation}
\Omega'_{i}=\frac{3}{a}\Omega_i\left[\left(\sum_{i} w_{i}
\Omega_{i}\right) - w_{i}\right]\, , \label{omegaprime-i}
\end{equation}
and obtain
\begin{equation}
S''_{H} \propto \frac{3}{a^2 H^2}\left[a\sum_i w'_i \Omega_i+ 6
\left(\sum_{i} w_{i} \Omega_{i}\right)^2+\sum_{i} w_{i} \Omega_{i}
(5-3w_{i})+2\right]\, .
\label{eqhor}
\end{equation}
As readily seen, $S''_{H}(a \rightarrow \infty) \rightarrow 0$, is
in agreement with the graph of $S_{H}$ of Fig.
\ref{fig:entropyhor}.
\\  \

\noindent From Eqs. (\ref{S'hor}) and (\ref{eqhor}) we infer  that
the apparent horizon of the UDF model of Ref. \cite{yang13} (and,
in general, of every reasonable UDF model) satisfies the second
law of thermodynamics. However, it would be too premature to jump
to the conclusion that this guarantees the fulfillment of the
second law for the Universe itself. It could happen that at some
stage of the expansion the said law would get violated by one or
more fluid components and a breakdown of the second law would be
induced. Nevertheless, given the multiplicative factor $k_{B} \,
\ell_{pl}^{-2}$ in the expression for $S_{H}$, it is natural to
expect that, indeed, the entropy of the horizon dominates over
that of every component. In fact, this is the case by a factor of
$18$ orders of magnitude in the present Universe \cite{egan2010}.
At any rate, it is safer to investigate the behavior of the two
first derivatives of the fluid components to check whether the
total entropy, given by the right-hand side of (\ref{Stotal}),
comply with the said law.
\\  \

\noindent The variation of the entropy  of the radiation fluid and
the unified fluid component follows from  Gibbs's law
\begin{equation}
T_{k} \, dS_{k} =\ d\left(\rho_{k}\, \frac{4\pi
\tilde{r}_{H}^{3}}{3}\right)\, + \, \ p_{k} \, d\left(\frac{4\pi
\tilde{r}_{H}^{3}}{3} \right)\, ,  \qquad (k = r,\, u)
\label{gibbs}
\end{equation}
where $T_{k}$ denotes the fluid temperature, which is always
positive definite. With the help of (\ref{Friedmann}) and
(\ref{Hprime}) it can be recast as
\begin{equation}
T_{k} \, S'_{k} =\frac{3 \Omega_{k} (1+w_{k})}{4 G a H}\left[1\, +
\, 3 \left( w_{r}\Omega_{r} \, + \,  w_{u}\Omega_{u}\right)
\right] \, . \label{fluidSL}
\end{equation}
Bearing in mind the expression for $w_{u}$ [Eq. (\ref{eos})] and
Fig. \ref{fig:densities}, one realizes that, except when the scale
factor is small, $S'_{r}$ and $S'_{u} \, $ are bound to be
negative.
\\ \

\noindent From  Gibbs equation and the condition that $\, dS_{k}$
be a differential, one obtains $\, d \ln T_{k}/d \ln a = -3 w_{k}
\, $. Consequently,
\begin{equation}
T_{r} = T_{r0}\, a^{-1}\, \quad {\rm and} \quad
T_u=T_{u0}\exp{\left[-3\int_{1}^{a} \frac{w_u(x)}{x} dx \right]}
\,. \label{tempevol}
\end{equation}

\noindent Because the baryon fluid behaves  essentially as dust,
its temperature vanishes and Gibbs's equation cannot be employed
to calculate the evolution of its entropy. Here we resort to the
procedure followed in \cite{grg1}. Consider  that every dust
particle contributes  to the entropy of this component by a given
bit, say $k_{B}$. Hence, within the apparent horizon we will have
$S_{b}=k_{B} N$, where $\, N = n (4 \pi/3) \tilde{r}_{H}^{3} \,$
denotes the number of particles there and $\, n = n_{0} \, a^{-3}
\, $ is the number density of dust particles. Thus, with the help
of (\ref{Hprime}) we get
\begin{equation}\label{dustentropy}
S'_{b} =\frac{2 \pi n_{0}  k_B}{a^{4} H^{3}}\left[1\, + \,
3\left(w_{r} \Omega_{r} + w_{u} \Omega_{u}\right)\right] \, .
\end{equation}
Again, it is apparent that from some scale factor onward $S'_{b}$
will be negative.
\\  \

\begin{figure}[htb]
\hspace{-2.5cm}
\begin{minipage}{0.3\textwidth}
\centering
 \includegraphics[width=8cm]{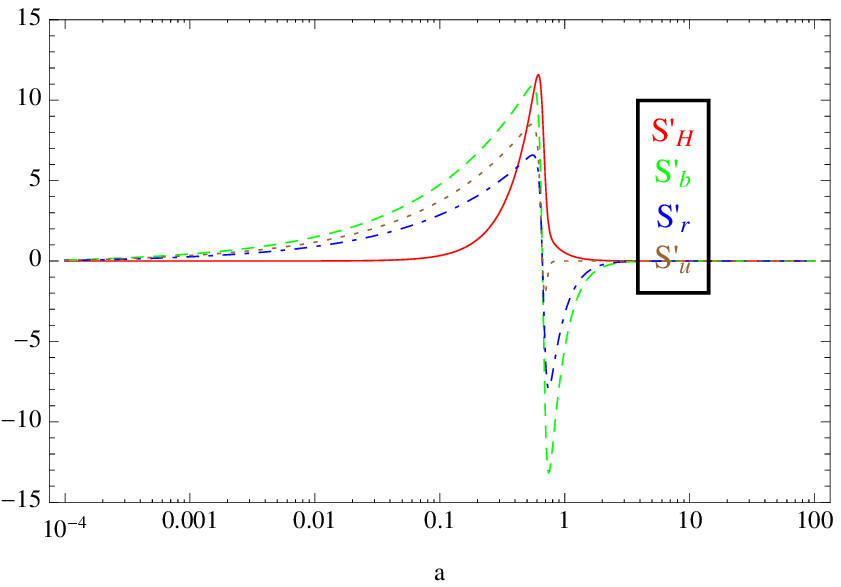}
 \end{minipage}
 \hspace{4 cm}
\begin{minipage}{0.3\textwidth}
\centering
 \includegraphics[width=8cm]{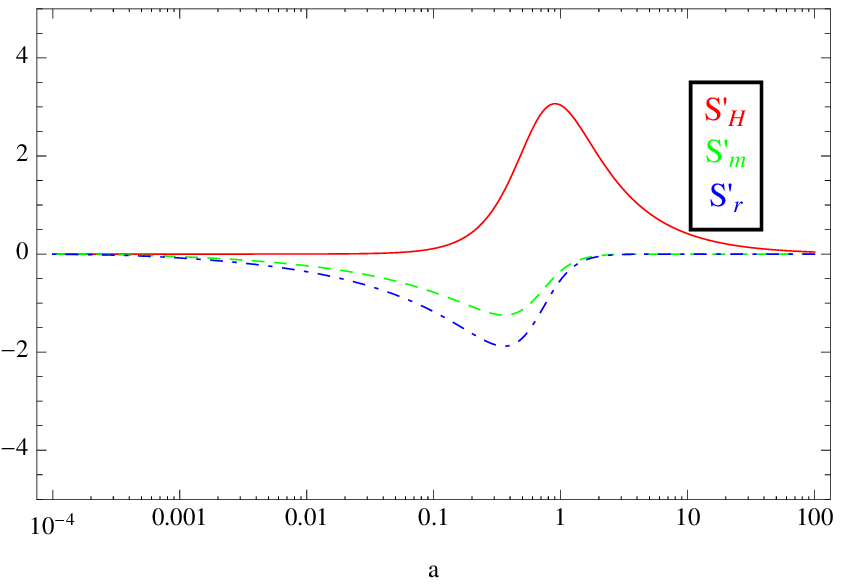}
  \end{minipage}
  \caption{{\small Left panel: Evolution of $dS/da$ for the apparent horizon and the
  energy components of the UDF model. The scales for $S'_{H}, S'_{u},
  S'_{r}$ and $S'_{b}$, should be multiplied by a factor of $10^{99}, 10^{89}, 10^{77}$,
  and $10^{50}$, respectively. Note that, $S'_{u}, S'_{r}$ and
  $S'_{b}$ becomes negative near the present epoch and remains so
  forever. However, $S'$ is positive all the same because $S'_{H}$
  dominates by a huge margin. In plotting the graphs we used the same values as in Fig.
  \ref{fig:densities}. Right panel: {\em Idem} for the $\Lambda$CDM model.
  The subscript $m$ stands for  the pressureless energy components, baryons plus cold
  dark matter. In plotting the graphs we used $\, \Omega_{r0} = 5 \times 10^{-5}$, $\,
  \Omega_{m0} = 0.27$ and $\, \Omega_{\Lambda 0} = 1- \Omega_{r0} -
  \Omega_{m0}$.}}
  \label{fig:fluidSL}
\end{figure}

\noindent Figure \ref{fig:fluidSL} depicts the evolution of
$dS_{i}/da \,$ ($i = H, u, r, b$) for the UDF model (left panel)
and the concordance $\Lambda$CDM model (right panel). The latter
is shown for the sake of comparison.

\begin{figure}[htb]
\hspace{-2.5cm}
\begin{minipage}{0.3\textwidth}
\centering
 \includegraphics[width=8cm]{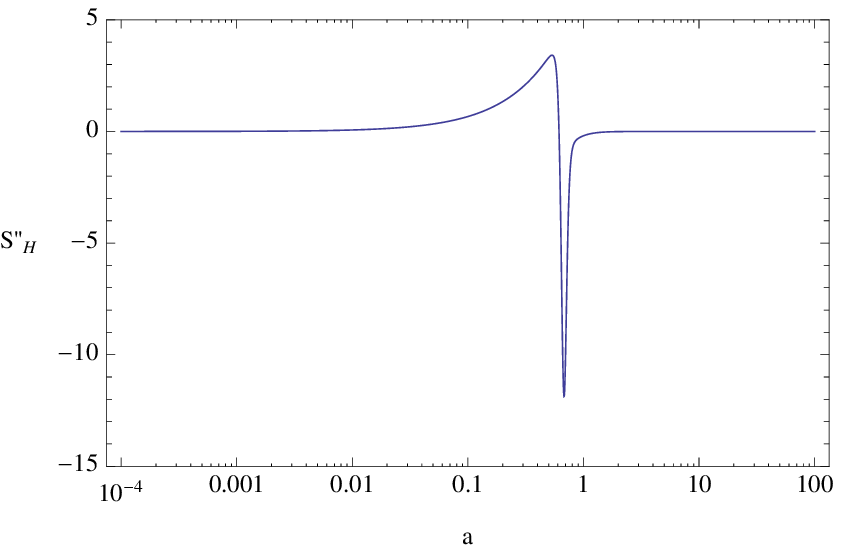}
 \end{minipage}
 \hspace{4 cm}
\begin{minipage}{0.3\textwidth}
\centering
 \includegraphics[width=8cm]{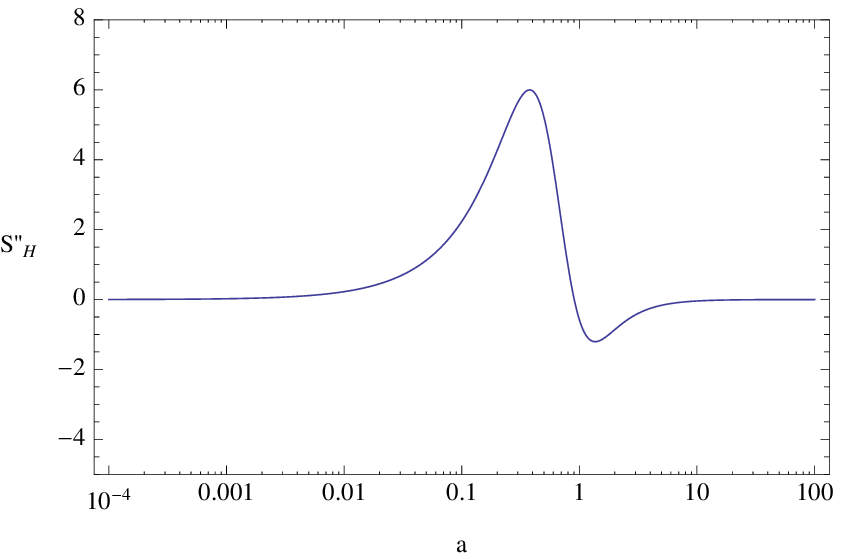}
  \end{minipage}
  \caption{{\small Left panel: $d^{2}S_{H}/da^{2}$ vs. the scale factor for the UDF model.
  In drawing the plot we used the values employed in Fig. \ref{fig:densities}.
  Right panel: The same but for the concordance $\Lambda$CDM model. In plotting
  this graph we used $\, \Omega_{r0} = 5 \times 10^{-5}$,
  $\, \Omega_{m0} = 0.27$, and $\, \Omega_{\Lambda 0} = 1- \Omega_{r0} -
  \Omega_{m0}$.}}
  \label{fig:S2primeH}
\end{figure}

\noindent Likewise, Fig. \ref{fig:S2primeH} depicts the evolution
of $S''_{H}$  for the UDF model (left panel) and the concordance
$\Lambda$CDM model (right panel). The corresponding second
derivatives of the fluid components of either model are not shown
because they fall by many magnitude orders below $S''_{H}$,
whereby their graphs  would essentially overlap the horizontal
axis.
\\ \

\noindent Given the overwhelming dominance of the entropy of the
horizon  (and its two first derivatives) over the entropies of the
fluid components, it follows that the second law of thermodynamics
is satisfied by the UDF model (as well as by the $\Lambda$CDM
model) thanks to the behavior of $S_{H}$. Put another way, the
fluid components  do not by themselves satisfy the aforementioned
law. If it were not by the horizon entropy neither model would
comply with it. Should it be so, one would conclude that either of
these models are unphysical or that the second law does not apply
to cosmological scales. However, the latter conclusion would be
hard to swallow in view, as mentioned above,  of the close link
between gravitation and thermodynamics \cite{jakob,steven,
jacobson,paddy}.

\begin{figure}[htb]
\centering
\includegraphics[width=10cm]{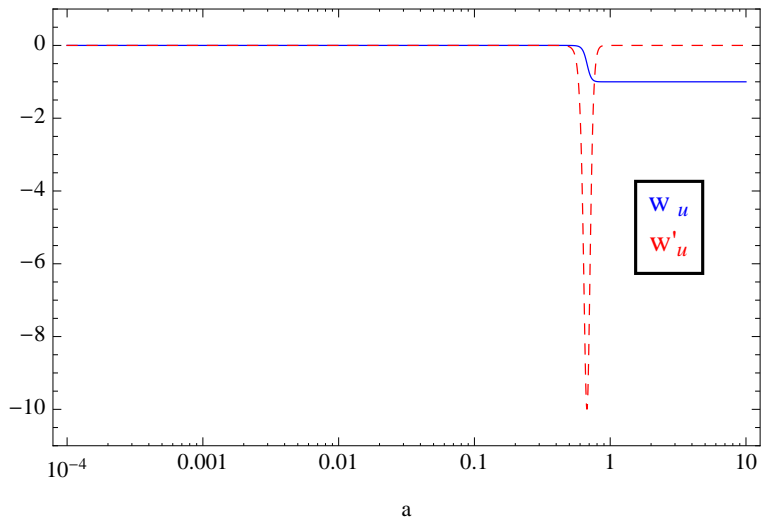}
\caption{{\small The EoS of the UDF (solid line) fluid and its
first derivative (dashed line) vs. the scale factor. In plotting
the graphs we used $\, a_{t} = 0.674\,$ and $\, \beta = 0.249$.}}
\label{fig:wwprime}
\end{figure}

\noindent The left panels of Figs. \ref{fig:fluidSL} and
\ref{fig:S2primeH} show a strong and sudden oscillation in the
entropy derivatives of the UDF model that starts well before the
transition scale factor $\, a_{t}$ is attained (i.e., while
$w_{u}$ still mimics the EoS of pressureless dark matter) and ends
up shortly after the present time. By contrast, as shown in the
right panels of these figures, the oscillation is much less severe
in the $\Lambda$CDM model (it is of a much smaller amplitude, and
grows and decays more slowly) and finishes much later (especially
in $S'_{H}$).
\\   \

\noindent This different behavior can be traced to the evolution
of $\, dw_{u}/da \,$ depicted in Fig. \ref{fig:wwprime}. The $\,
w'_{u} \, $ big dip, centered about $\, a_{t} \,$, accounts for
the quick variation of the first derivative of the entropies, and
the second derivative of $S_{H}$. By contrast, in the $\Lambda$CDM
model (which has $w_{\Lambda} = -1\, $ and $ \, w_{m} = 0\, $ at
any scale factor), the said variations are necessarily much
softer.
\\  \

\noindent In summary, on the one hand, the UDF class of models
fulfill the second law of thermodynamics, i.e., its total entropy
(that of the horizon plus matter and fields inside it is never
decreasing and it tends to a maximum as $a \rightarrow \infty$).
On the other hand, owing to the abrupt behavior of the EoS, the
first and second derivatives of the entropy present a rather
peculiar, sharp oscillation that casts doubts on the soundness of
this class of models.
\\ \

\noindent NR thanks the ``Agenzia Spaziale Italiana" (ASI) for
partial support through Contract No. I/034/12/0. This work was
partially supported the ``Ministerio de Econom\'{\i}a y
Competitividad, Direcci\'{o}n General de Investigaci\'{o}n
Cient\'{\i}fica y T\'{e}cnica", Grant No. FIS2012-32099.


\end{document}